\documentclass[10pt,twocolumn,letterpaper]{article}

\usepackage[pagenumbers]{iccv} 

\usepackage{multirow}
\usepackage{comment}
\usepackage{animate}
\usepackage{colortbl, array, xcolor}
\usepackage{listings}
\newcommand{\figref}[1]{Figure~\ref{#1}}

\newcommand{\ours}[1]{MG-Gen}
\newcommand{\Ours}[1]{MG-Gen}

\definecolor{iccvblue}{rgb}{0.21,0.49,0.74}
\usepackage[pagebackref,breaklinks,colorlinks,allcolors=iccvblue]{hyperref}

\def\paperID{*****} 
\def\confName{ICCV}
\def\confYear{2025}

\title{MG-Gen: Single Image to Motion Graphics Generation}
\author{
    Takahiro Shirakawa \and 
    Tomoyuki Suzuki \and 
    Takuto Narumoto \and 
    Daichi Haraguchi \and
    CyberAgent\\
    {\tt\small \{shirakawa\_takahiro, suzuki\_tomoyuki, narumoto\_takuto, haraguchi\_daichi\_xa\}@cyberagent.co.jp}
}

\begin{document}
\twocolumn[{
\renewcommand\twocolumn[1][]{#1}%
\maketitle
\begin{center}
    \centering
    \captionsetup{type=figure}
    \includegraphics[width=.975\textwidth]{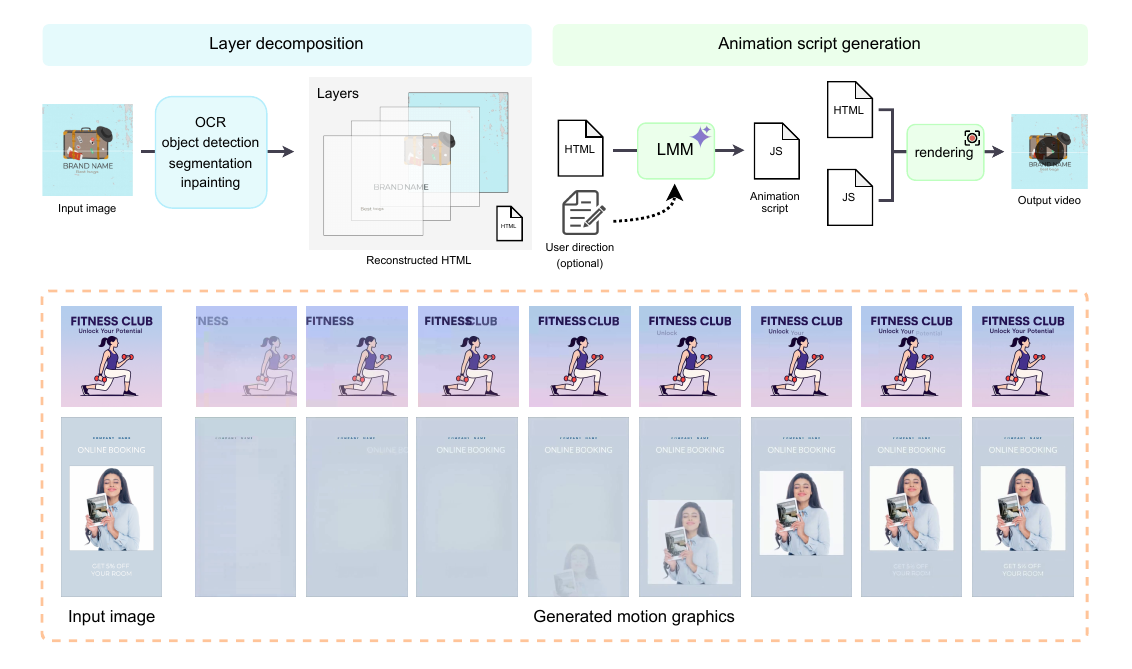}\\[-2mm]
    \caption{Overview of our single raster image to motion graphics generation framework (\ours{}). Generated motion graphics have dynamic motion while preserving text legibility and object geometry.}
    \label{fig:teaser}
\end{center}
}]

\begin{abstract}
\looseness=-1
We introduce \ours{}, a framework that generates motion graphics directly from a single raster image. \ours{} decompose a single raster image into layered structures represented as HTML, generate animation scripts for each layer, and then render them into a video.
Experiments confirm \ours{} generates dynamic motion graphics while preserving text readability and fidelity to the input conditions, whereas state-of-the-art image-to-video generation methods struggle with them.
The code is available at \url{https://github.com/CyberAgentAILab/MG-GEN}
\end{abstract}

\section{Introduction} \label{sec:intro}
Motion graphics are videos that bring static graphic-design elements--such as text, shapes, and icons--to life through animation, conveying concise messages and adding visual impact. As short-form video content grows on social media, motion graphics demand has surged for advertisements, music videos, and personal content. However, motion graphics production is complex, requiring advanced techniques, precise motion control, and careful effect selection.

Motion graphics differ from natural videos in two key respects. First, because they animate design assets like text and icons, legibility must be maintained throughout the motion, while natural videos focus instead on photorealism and narrative realism. 
Second, the shapes and colors of these assets must stay geometrically stable to protect brand integrity, whereas camera footage can tolerate minor distortions. 
To meet these requirements, we formulate a new task: generating motion graphics from a single raster image.
This task involves not only creating visually appealing animations, but also preserving the design semantics of each component—such as text, icons, and layout—and enabling precise, element-wise motion control. 
It especially demands content awareness and eye-catching animation tailored to each visual element.

We experimentally confirmed that general image-to-video methods~\cite{blattmann2023stable,veo,gen3,ray2,wan21,hunyuan} generate high-quality natural videos but struggle with motion graphics generation. 
This is because these models generate entire raster videos directly, rather than synthesizing animation for individual objects or layers. 
As a result, they often fail to maintain the precise shape and spatial consistency of design elements, such as text, logos, or product pictures.
Moreover, these methods lack precise control over which elements move and how they move. For example, users cannot explicitly specify that a particular text should fade in from left to right, making it difficult to realize intentional and structured animations.
Aside from generating raster videos directly, existing motion graphics generation approach~\cite{liu2025logomotion} animates each layer individually by generating an animation script specifying animation parameters of them. This approach is reasonable and generates high quality motion graphics, preserving content consistency and generating dynamic motion but it requires already organized as layer data (e.g., PDF) which is rarely available for the system input.

To address these limitations, we propose \ours{}, the first end-to-end framework that transforms a single raster image into a motion graphics video. 
Optionally, users can also provide motion instructions as user direction, and \ours{} generates the video to match these directions.
\ours{} consists of two steps: layer decomposition, which transforms a single raster image into layered hierarchical structure data, and animation script generation, which synthesizes executable animation script for each decomposed layer.
In layer decomposition, we combine various task specific models such as OCR, object detection, segmentation, and inpainting models, yielding clean layers. The decomposed layers are reconstructed as HTML data with a hierarchical structure.
In animation script generation, we leverage the performance of a large multimodal model (LMM) to generate an executable animation script that synthesizes the layer-level animation based on previous work~\cite{liu2025logomotion}.
LMMs can generate complex but error-free script thanks to their high coding proficiency. Furthermore, their superior reasoning abilities enable them to formulate effective animation plans and precisely interpret and incorporate user directions.
Finally, we combine the generated animation script with HTML data, then render it to produce the video.

In our experiments, we first qualitatively observe that \ours{} can generate motion graphics with dynamic text motion, high readability, and reduced object distortion.
We further compare the results generated by \ours{} with state-of-the-art general image-to-video generation baselines.
Based on the results from both GPT-based and human evaluations, we find that \ours{} outperforms the baselines in terms of dynamic text motion, while also achieving superior readability and fidelity to the input image.

Our main contributions are summarized as follows:
\begin{itemize}
  \item We introduce the new task of \emph{motion graphics generation from a single raster image}, clarifying the constraints that set it apart from conventional image-to-video synthesis.
  \item We propose \ours{}, a two-stage framework that 
  (i) decomposes the input image into a structured, layered HTML, and
  (ii) generates layer-wise animation script with an LMM, achieving dynamic motion while preserving text legibility and object geometry.
  \item Extensive experiments demonstrate that \ours{} outperforms state-of-the-art image-to-video baselines on both human evaluation and GPT-based metrics.
\end{itemize}

\section{Related Work}  \label{sec:related-work}
\subsection{Image-to-Video Generation}
Image-to-video generation is gaining attention due to recent advancements in video generation methods~\cite{singer2022make,ho2022imagen,ho2022video,jiang2024videobooth,blattmann2023stable,shi2024motion,wang2024videocomposer,ni2023conditional,hu2022make,ren2024consisti2v,gen3,veo,kling}.
Several approaches accept not only images (and prompts) but also additional conditions, such as motion trajectory~\cite{shi2024motion} and image depth~\cite{wang2024videocomposer}, among others.
Recent advancements in image-to-video generation, such as Ray2~\cite{ray2}, Hunyuan~\cite{hunyuan}, and Wan2.1~\cite{wan21}, have shown promising results in producing high-quality videos with substantial frame-to-frame consistency.
While the field of video generation has seen significant advancements, few studies have specifically focused on generating videos within the specialized domain of motion graphics. This study aims to directly address this notable research gap by enabling the motion graphics generation.

\subsection{Animated Text Generation}
Animated text with visual effects is widely used across various fields, including movies, lyric videos, advertisements, and social media. However, creating these videos is time-consuming, prompting the development of support tools~\cite{forlizzi2003kinedit,kato2015textalive,xie2023emordle} and automated generation methods~\cite{xie2023wakey,park2024kinety,shin2025generating} to reduce the workload.
Xie et al.~\cite{xie2023wakey} proposed a method to transfer animations from GIFs to vector-based text. 
Park et al.~\cite{park2024kinety} introduced a diffusion-based approach for generating kinetic typography videos in raster space based on user instructions.
Shin et al.~\cite{shin2025generating} proposed an animated layout generation to animate texts for advertising videos.
However, these approaches exclusively focus on animating text, whereas motion graphics often require simultaneous animation of both textual and non-textual (object) elements.

\subsection{Video Generation with LLMs and LMMs}
Large language models (LLMs) and large multimodal models (LMMs) have recently been applied to video generation in two primary ways.
One approach uses them as high-level planners that generate prompts or conditions for downstream generative models~\cite{huang2024free-bloom,lian2023llm-grounded,lin2023videodirectorgpt}.
Another direction is code-based video generation~\cite{liu2025logomotion,tseng2024keyframer,lv2024gpt4motion,generativemanim}, which leverages the code generation capabilities of LLMs and LMMs to produce animation scripts directly.

LogoMotion~\cite{liu2025logomotion} and Keyframer~\cite{tseng2024keyframer} animate static PDF and SVG files by generating JavaScript and CSS animation code, using an LMM and an LLM, respectively.
Although these animations look more natural than those produced by general image-to-video methods, they are limited to accepting PDF or SVG inputs, which are not as widely available as raster images.
These limitations motivate our work, which aims to enable code-based animation generation from raster images using an LMM.

\section{Method} \label{sec:method}
\subsection{Overview}
Given an input raster image, \ours{} generates motion graphics that accurately display dynamic animations while maintaining text readability and consistency with elements in the input image, such as text, logos, and product pictures.
As shown in \figref{fig:teaser}, \ours{} decomposes an input image into layered hierarchical structure data represented as HTML and generates animation script for the HTML by using an LMM. The HTML and generated script are rendered into video data.
\ours{} incorporates OCR~\cite{documentai}, object detection~\cite{wang2024yolov9}, object segmentation~\cite{ye2024hi-sam,ravi2024sam}, and image inpainting~\cite{suvorov2022resolution} for layer decomposition of the input image, and then reconstructs the decomposed layers as HTML data that visually replicates the input image.
It also utilizes the LMM, which has shown strong performance in code understanding and code generation tasks, following the approach of prior work~\cite{liu2025logomotion} for generating executable JavaScript animation scripts. 

\begin{figure}[t]
    \centering
    \includegraphics[width=\linewidth]{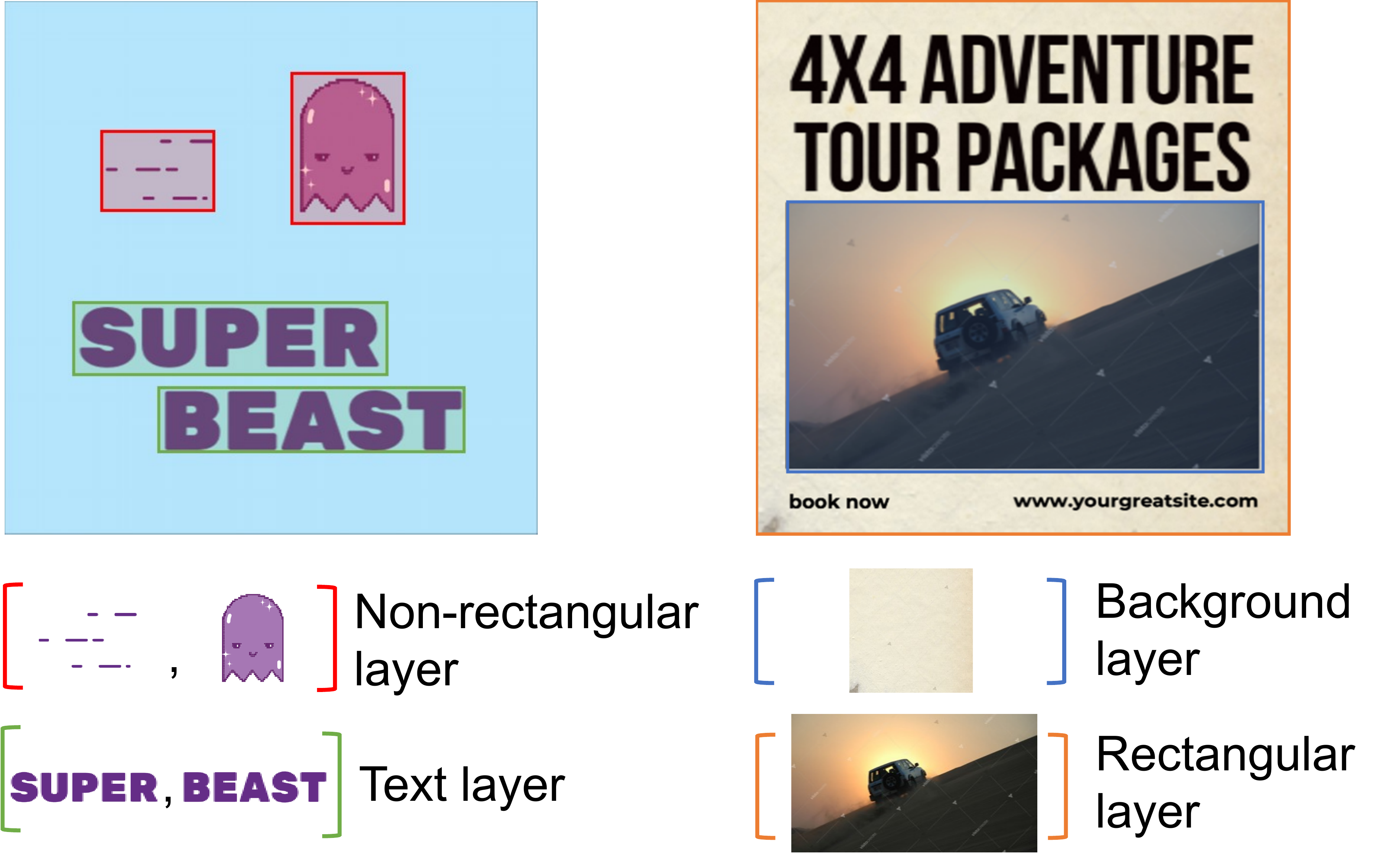}\\[-2mm]
    \caption{Examples of each layer.}
    \label{fig:layer-detail}
\end{figure}

\begin{figure*}[t]
    \centering
    \includegraphics[width=0.95\linewidth]{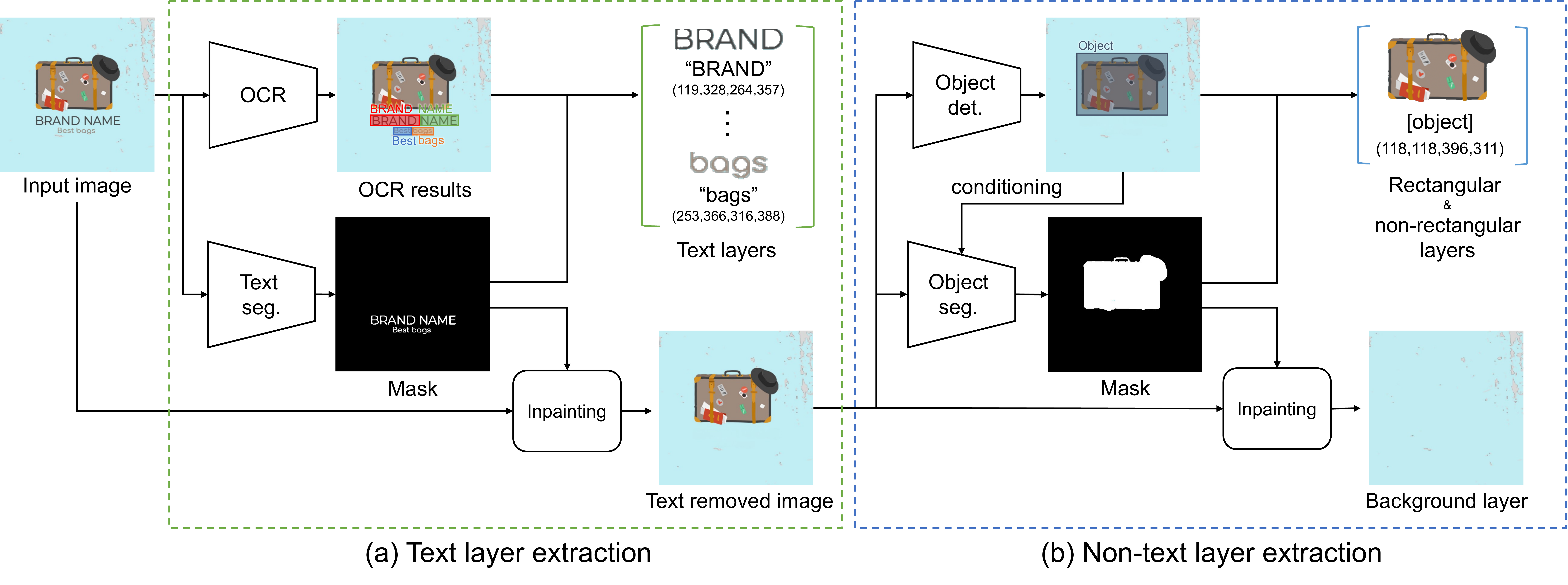}\\[-3mm]
    \caption{Overview of layer decomposition, which consists of two steps: (a) text layer extraction and (b) non-text layer extraction. For OCR, segmentation, and inpainting models, we utilize publicly available pre-trained models, with fine-tuning exclusively applied to the object detection model.}
    \label{fig:layer-decomposition}
\end{figure*}

\subsection{Layer Decomposition} \label{sec:method/layerdec}
We illustrate the layer decomposition process of \ours{} in \figref{fig:layer-decomposition}. 
First, only text layers are extracted using OCR and text stroke segmentation models, following the graphic design principle that text typically appears on top.
Then, we extract the remaining layers, which include illustrations, decorative elements, and photographs, from the text-removed image, using object detection and segmentation models, followed by image inpainting to extract a separate background layer behind these layers.

\paragraph{Text Layer Decomposition}
In this process, we extract only textual elements from the input image, with each word isolated as a separate image layer, as shown in \figref{fig:layer-decomposition} (a).
We first obtain the text contents and their bounding boxes from the input image using OCR model, DocumentAI~\cite{documentai}. Also, we obtain a text stroke mask using Hi-SAM~\cite{ye2024hi-sam}, a specialized segmentation model for text.
Then, we create text image layers by grouping the text stroke mask based on the bounding boxes obtained from OCR and using each grouped stroke mask as an alpha channel for the corresponding text layer.
Finally, we remove the text from the input image using an image inpainting model, LaMa~\cite{suvorov2022resolution}. This text-removed image is used for the subsequent non-text layer extraction process.

\paragraph{Non-text Layer Decomposition}
In this process, we extract remaining non-text layers from the text-removed image, as shown in \figref{fig:layer-decomposition} (b).
We process the layers separately as rectangular and non-rectangular ones (\figref{fig:layer-detail}), since the former can be easily extracted using bounding boxes, whereas the latter requires more complex segmentation. Most rectangular layers consist of product or person photographs, while non-rectangular layers are typically vector illustrations or decorative elements. 

We first obtain the bounding boxes and types (rectangular or non-rectangular) of the non-text layers in the text-removed image, except the background layer, using object detection model, YOLOv11~\cite{wang2024yolov9}. For the detection of these layers, we train the model on the Crello dataset~\cite{yamaguchi2021canvasvae}, which comprises graphic designs with a layered structure. To construct the training dataset, we render images without text, generate bounding boxes for non-text layers, and label each with a type based on criteria such as whether it has a rectangular shape. We finally obtain 18,864 training images and 1,772 validation images, which we use to train the detector.

After detecting the layers, we apply SAM2~\cite{ravi2024sam} only to the non-rectangular ones, conditioned on the bounding boxes, to obtain their segmentation masks. For rectangular layers, we directly use the detected bounding boxes as masks, as their shapes are inherently rectangular.
Similar to the text layer, each element is cropped based on its bounding box, and its estimated mask serves as an alpha channel to extract it as a distinct layer.

Finally, we extract the background layer by inpainting the regions behind the extracted all layers except background layer using LaMa.

\paragraph{HTML Reconstruction}
We compose each text, non-text, and background layers into HTML using \texttt{<img>} tags based on the bounding box information extracted by OCR and layer detection. Each \texttt{<img>} tag is enriched with attributes including a unique ID, a layer type, and a caption for the image layer generated by the LMM (or OCR results for text layers). 
Here, the layer type is one of the four decomposed in the previous phases: text, rectangular, non-rectangular, and background.
This information aids in producing more effective animations when generating the animation script. Note that each text layer is treated as a discrete image rather than in vector format.

\begin{figure*}[t]
    \centering
    \includegraphics[width=\linewidth]{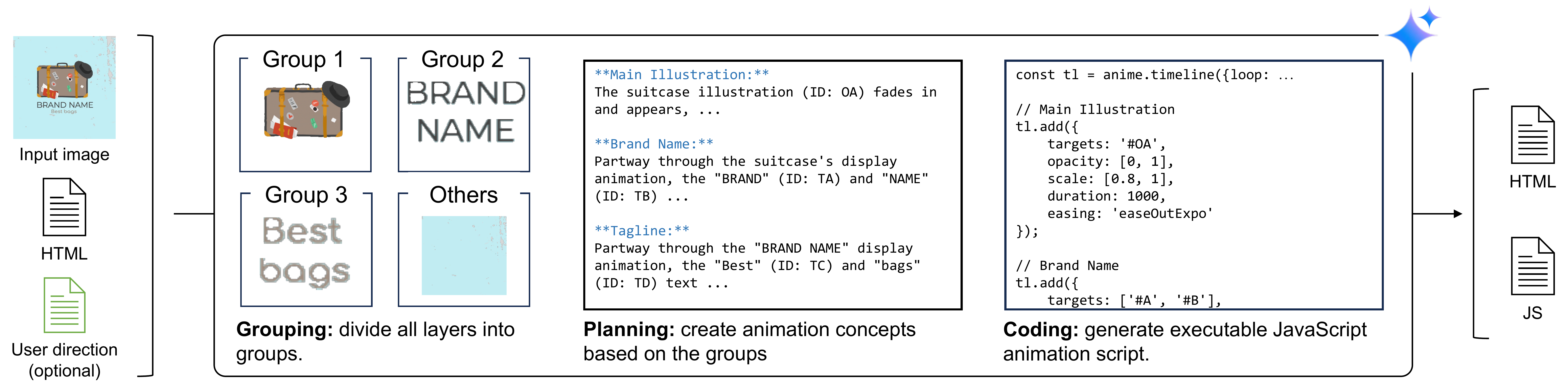}\\[-2mm]
    \caption{Overview of animation script generation. The user direction (motion instruction) is an optional prompt to specify the motion of each layer.
    }
    \label{fig:animation-generation}
\end{figure*}

\subsection{Animation Script Generation}
We illustrate animation script generation process of \ours{} in \figref{fig:animation-generation}. Inspired by prior script-based animation generation~\cite{liu2025logomotion}, we use an LMM to generate an executable JavaScript animation script. Animation script generation pipeline consist of three main process: layer grouping, animation planning, and script coding. 

\paragraph{Grouping}
We first classify each layer of the input HTML into distinct groups or clusters, which the LMM then reads to understand the HTML's structure and form clusters suitable for animation generation. It enables effective animation script generation for each group. 

\paragraph{Planning}
We then generate animation plan based on the input HTML and layer group information. The LMM outputs a plan detailing what animation to generate for each group and how to connect these groups into a cohesive animation.
When provided with a user direction, it outputs an animation plan that incorporates the user's intent.

\paragraph{Coding}
We finally generate executable JavaScript animation script based on the input HTML and animation plan. Script is built upon Anime.js~\footnote{\url{https://animejs.com/}}, a lightweight JavaScript library. It works with any CSS properties in HTML such as position, color, and scale, and it can generate diverse animations.

\subsubsection{Detailed Implementation of LMM}
We utilized Gemini-2.5-Pro~\cite{gemini, team2025gemini} as the LMM for animation script generation due to its high code comprehension and generation capabilities, coupled with its advanced reasoning abilities. 
The processes are executed in a chat-based interaction. We give the input image used for layer decomposition and prompt instructions provided below.
We first obtain group information by the prompt as follows:
\begin{lstlisting}[]
Please divide all layers into several animation groups considering the layout and content of each layer based on given image 1 and html. The HTML below represents the above image in HTML format.
```[#HTML]```
\end{lstlisting}
Then, we generate an animation plan based on the HTML data, user direction as optional, and pre-generated group information by the prompt as follows: 
\begin{lstlisting}[]
Given image 1 shows the thumbnail of an input image for which we want to generate animation. Also, the HTML represents the image in HTML format.

Animation user instruction is given. if empty, create animation idea.
```[#USER]```

Please generate an animation plan to make the layer decomposed image objects and text appear on screen, taking into account the layout and each content of the image shown above. The IDs and contents of the layers we want to animate are as follows:
```[#LAYER]```
\end{lstlisting}
Finally, we generate an animation script based on the animation plan by the prompt as follows.
\begin{lstlisting}[]
Please generate an animation script using anime.js to implement these animations. The animation code should be written as follows:
- Use a single `anime.timeline` and add animations using `.add`
- Apply one animation per layer
- Coordinate movements of animations within the same group
- Configure the timeline settings with `loop=false` to prevent looping and `autoplay=true` to start the animation automatically.
\end{lstlisting}
During prompt input, \texttt{[\#HTML]}, \texttt{[\#USER]}, and \texttt{[\#LAYER]} are also replaced with actual data, such as the HTML source code, user motion instructions, and pre-generated group information in JSON format.
We optionally accept motion description prompt as user direction to specify the motion of each layer when generating the animation script. Note that this is an optional user operation to control the generated animation, and even without it, an appropriate animation plan can be generated. 
Further detailed prompt instructions can be found on our GitHub.

\subsubsection{Video Generation through Rendering}
Once the animation script is generated, we obtain a video file by rendering HTML and JavaScript files. We use Playwright~\footnote{\url{https://github.com/microsoft/playwright}} to automate this process. Animations are recorded at 25 fps by default to ensure smooth playback.

\begin{figure*}[t]
    \centering
    \includegraphics[width=\linewidth]{figs/results_comparison_v5_a.pdf}\\[-3mm]
   \caption{Motion graphics generated by \Ours{} without explicit motion instruction prompts.}
    \label{fig:results_comparison_a}
\end{figure*}

\begin{figure*}[t]
    \centering
    \includegraphics[width=\linewidth]{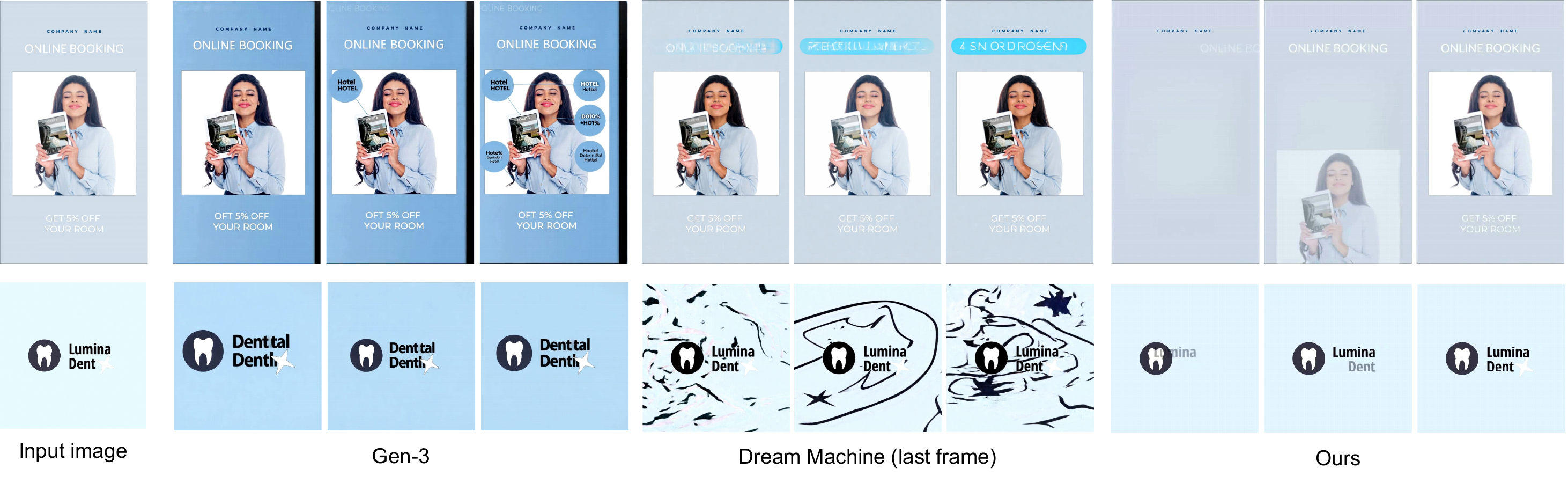}\\[-3mm]
   \caption{Visual comparisons of generated motion graphics by Ray2~\cite{ray2}, Wan2.1~\cite{wan21}, Hunyuan~\cite{hunyuan} and \Ours{} with the same motion instruction prompts.}
    \label{fig:results_comparison_b}
\end{figure*}

\section{Experiments}
\subsection{Experimental Settings}
\paragraph{Evaluation Dataset}
We use the Crello Animation dataset~\cite{suzuki2024fast}, which consists of motion graphics videos. We conduct validation for each method using the validation set and report the evaluation results on the test set. Note that this dataset is different from the Crello dataset~\cite{yamaguchi2021canvasvae} used to train the object detection model described in Section~\ref{sec:method}.
We generate motion description text from each video using GPT-4o~\cite{gpt4o}. 
The motion description serves as a motion instruction input of both \ours{} and comparative methods, replacing the user direction during the experiments.
Also, we extract the first and last frames from each video. We use the last frames as thumbnail images, which contain all elements of the entire frames of each video. 

\paragraph{Comparative Methods}
As there are no existing methods specifically designed for motion graphics generation from a single raster image, we compare our approach against three state-of-the-art general image-to-video generation baselines: Ray2~\cite{ray2}, Wan2.1~\cite{wan21}, and Hunyuan~\cite{hunyuan}.
We use publicly available web services for Ray2 and Wan2.1, while we ran Hunyuan locally to generate videos. By applying the LoRA model \footnote{\url{https://github.com/dashtoon/hunyuan-video-keyframe-control-lora}} for Hunyuan, we have enabled generation from two input images: the first and last frames. 
Note that \ours{} only takes the thumbnail (last frame) as image condition. 
For a fair comparison, we also provide each baseline with the same user-direction prompt, because without textual guidance these methods struggle to produce motion-graphics-like results.

\subsection{Qualitative Evaluation}
\figref{fig:results_comparison_a} shows the generated motion graphics by \Ours{}. In these examples, we don't use the user direction prompt (motion instruction) as input for \ours{}.
The four examples demonstrate that optimal motion graphics can be generated, as the layer decomposition appropriately extracts elements from images. \ours{} generates videos where text and illustrations move dynamically, all while maintaining text readability and consistency with the input image. 
Also, these videos are generated solely from a thumbnail image input, without explicit user motion instructions, demonstrating that our grouping, planning, and coding pipeline effectively interprets the input image and HTML data, propose optimal animation plans and generate corresponding scripts.
In the first row example, \Ours{} extracts elements including the text ``HAPPY VALENTINES DAY,'' ``I'd never seen love before I met you,'' a heart illustration, and a green frame object from the input image. 
The generated animation sequence shows ``HAPPY VALENTINES DAY'' sliding in from both the left and right sides, followed by the heart illustration popping up, and finally, the green frame appearing to display the text.
In the second row example, texts pop up from both left and right sides while the Earth appears from below, expanding in size, creating a dynamic animation sequence.
In the third row example, the animation shows a part of the logo appearing with a rotating motion to capture the viewer's attention, followed by the text ``CAR SERVICE'' sliding in from the top, and finally, text message from both sides simultaneously converging toward the center.
In the fourth row example, an illustration of a female model appears first, followed by text sliding in from the left.
\ours{} can generate dynamic motion graphics from a single image while generating dynamic text motion with readability and  preserving  the shapes of content elements.

\subsection{Visual Comparisons to State-of-the-Art Image-to-Video Methods}\label{sec:compara}
\figref{fig:results_comparison_b} shows visual comparisons of generated motion graphics by \Ours{} and state-of-the-art methods. Here, we use the same user direction prompt (motion instruction) as input for all methods such as 
\begin{lstlisting}[]
The text 'Local BICYCLE RENTAL' gradually appears at the top center of the frame, growing in size and becoming more prominent. Simultaneously, the phrase 'RIDE IN STYLE & COMFORT' emerges from the bottom left corner, moving towards the center. A yellow circular design element appears from the bottom left, expanding and rotating slightly as it moves into position. The background remains static, showcasing two people cycling on a wooden pier extending into a serene lake.
\end{lstlisting}
and
\begin{lstlisting}[]
Start with a solid orange background. Gradually reveal the top portion of a food image sliding down from the top. As the image settles, introduce the text 'Today's Special Menu' from the left, sliding in smoothly. Follow with the text 'Delivered fresh to your door' appearing below, sliding in from the left. Finally, add the text 'Order Now' and a phone number '+123-4567-890' sliding in from the right, completing the animation with all elements in place.
\end{lstlisting}

Compared to the others, \ours{} generates animations where text, photos, and objects move actively, enhancing their eye-catching effect while remaining faithful to the input thumbnail image.
In the upper examples, it accurately displays text while incorporating dynamic text motions.  The text "Local BICYCLE RENTAL" first appeared rhythmically on screen, followed by "RIDE IN STYLE AND COMFORT" sliding in from left to right, accompanied by the object designed to evoke a tire that enlarged it.
In the bottom examples, it features a natural sequence animation where the sushi picture slides in from the top, followed by text elements sliding in from both the left and right.

On the other hand, comparison methods frequently generate videos that include objects irrelevant to the thumbnail image or meaningless text. They also tend to feature unnatural motion graphics, such as awkward movements.
Ray2 often generates irrelevant objects and meaningless text within the video that are not present in the input thumbnail image. In the upper example, objects and text collapse into noise, while in the lower example, numerous text elements irrelevant to the thumbnail image appear, resulting in unnatural videos.
Wan2.1 often generates unnatural motion graphics videos, characterized by unnatural text and object movements. In the upper example, the text merely floats around in a small size, and in the lower example, the phone number changes randomly like a slot machine; both result in unnatural videos.
Hunyuan also often generates unnatural motion graphics videos, exhibiting abrupt screen darkening or sudden changes in background color.
In the upper example, the screen's color abruptly darkens, while in the lower example, the background suddenly changes from black to orange, resulting in an unnatural video.

\subsection{Human Evaluation}
Given the absence of established evaluation protocols for this novel task, we conduct a user study focusing on three key dimensions: input consistency, prompt consistency, and dynamic text motion with readability.
We compare \Ours{} with the latest video generation models introduced in Section~\ref{sec:compara}, using the same input conditions for all methods to ensure fairness.
Participants rate the videos on the five-point Likert scale.
The specific explanatory text for each aspect was presented to the participants as follows:
\begin{itemize}
    \item {\it ``The thumbnail image and the video are consistent (no unnecessary objects or text appear)'' (input consistency)}
    \item {\it ``The instruction and the video are consistent (the instruction is sufficiently reflected, and there are no unnecessary animations)'' (prompt consistency)}
    \item {\it ``The text includes sufficient animation while remaining readable'' (text motion)} 
\end{itemize} 
To ensure response quality, we include a validation question that requires a ``no'' response. All participants passed this check.
Due to the intensive nature of human evaluation and the need to maintain high annotation quality, we randomly sample 20 input images from Crello Animation~\cite{suzuki2024fast} and divide them into two sets. Each set is evaluated by 10 unique participants, with no overlap. For each input image, videos generated by different methods are presented in randomized order, and participants provide scores on all three aspects. As a result, we collect evaluation responses from 10 participants per video per aspect.

\begin{figure}[t]
    \centering
    \includegraphics[width=.95\linewidth]{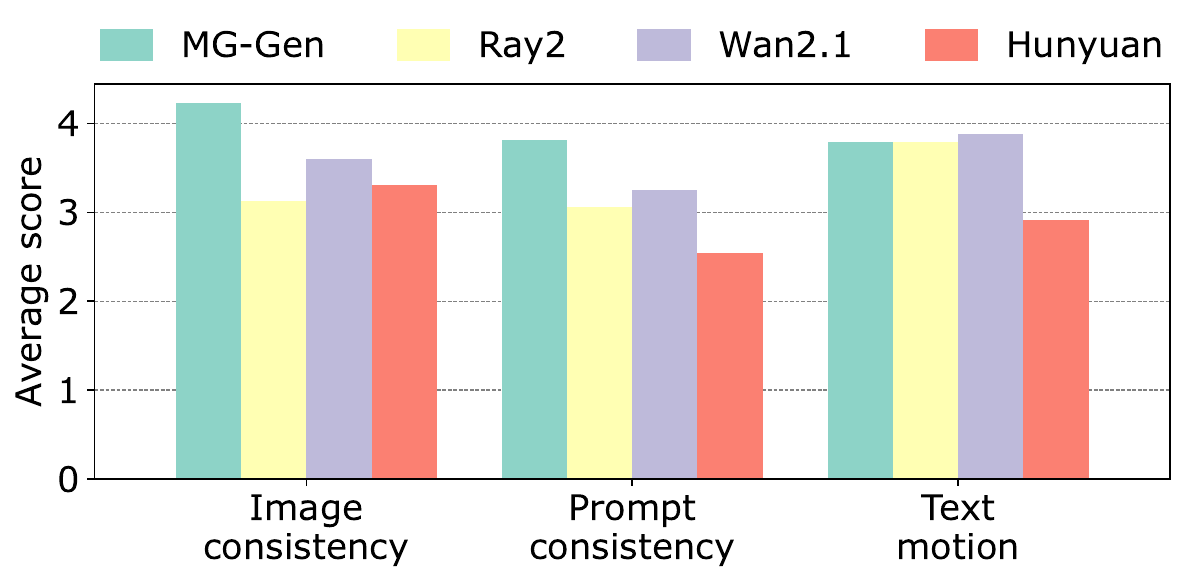}\\[-3mm]
    \caption{The results of user study.}
    \label{fig:user-study}
\end{figure}

We show the result of the user study in Figure~\ref{fig:user-study}. 
\Ours{} significantly outperforms the other three methods in terms of input consistency and prompt consistency.
Although Ray2 and Wan2.1 achieve comparable scores in text motion due to their ability to animate text dynamically, the animated text is often meaningless or misaligned with user intention, as indicated by lower scores in input and prompt consistency and the qualitative analysis in Section~\ref{sec:compara}.
These limitations of Ray2 and Wan2.1 are evaluated more strictly in the GPT-based evaluation, as discussed in the following section.
We emphasize that \Ours{} can generate natural animations while ensuring consistency with the input image and the motion instruction.

\begin{figure}[t]
    \centering
    \includegraphics[width=.95\linewidth]{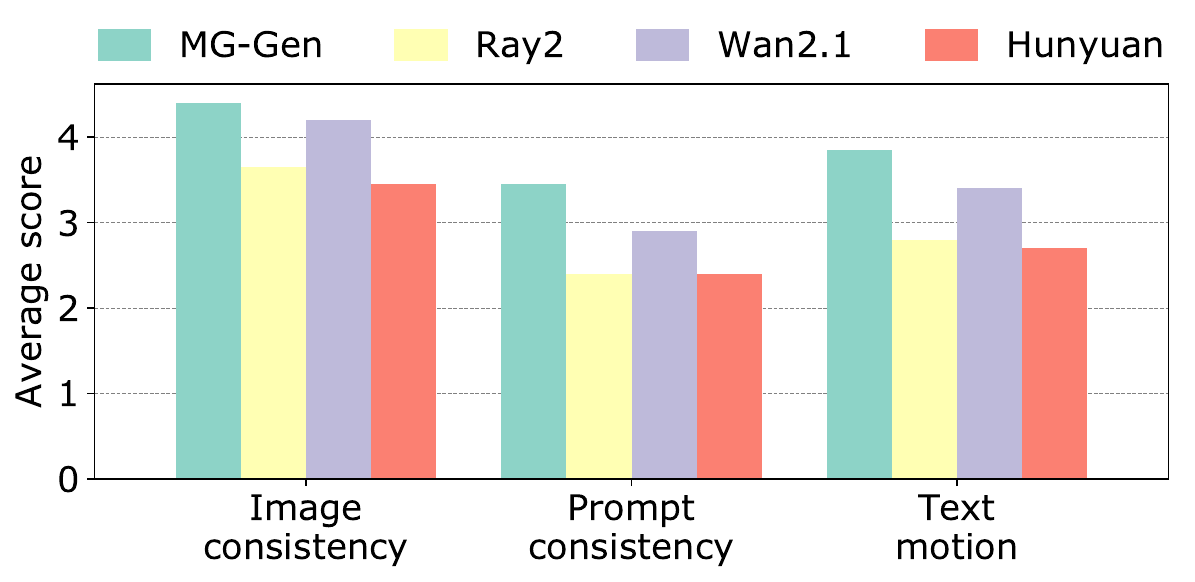}\\[-3mm]
    \caption{The results of GPT-based evaluation.}
    \label{fig:gpt-eval}
\end{figure}

\subsection{GPT-based Evaluation}
To complement the user study and reinforce the robustness of our evaluation, we also perform a GPT-based Evaluation using the exact same protocol. 
GPT is prompted to rate each video along the same three dimensions as the human participants, enabling consistent and scalable comparison across all methods.
For the evaluation, we employ GPT-o3~\cite{gpto3}.
Several GPT series can accept up to 50 images per a estimation.\footnote{GPT-4o/4.1 allow up to 50 input images; GPT-o3’s limit isn’t specified, so we apply the same.}
Therefore, we uniformly sampled 49 frames per video and use these frames and a thumbnail image, totally 50 images, as an input image.
The evaluation aspects and evaluated videos are the same as the human evaluation.
We also calculate the Pearson correlation coefficient between human evaluation scores and GPT-based evaluation scores.
The correlation coefficients were 0.520 for image consistency, 0.531 for prompt consistency, and 0.402 for text motion.
The correlations observed across all aspects are statistically significant ($p < 0.01$), indicating a consistent and meaningful relationship.

As shown in Figure~\ref{fig:gpt-eval}, \ours{} outperforms the other three methods in all aspects.
Different from human evaluation, \ours{} outperforms in text motion.
GPT consistently assigns lower scores to videos with collapsed texts.
This pattern was notably observed in Wan 2.1 and Ray2.
Consequently, the model's scores may not fully align with human judgments.
In contrast to these methods, \ours{} maintains text stability even under moderate motion, without exhibiting severe collapse.

Based on the results from both human and GPT-based evaluation, the proposed method demonstrates superiority in the following aspects.
\begin{itemize}
    \item \textbf{Faithfulness of user intent}: \ours{} consistently outperforms the baselines in input image and prompt consistency. This is important for generating motion graphics that align with the user's intent.
    \item \textbf{Satisfaction of requirements of motion graphics}: \ours{} is capable of generating motion graphics with sufficient text motion while avoiding severe text collapse. This is an important requirement for motion graphics, as it ensures that the message is immediately recognizable and brand consistency is preserved. In addition, high input image consistency suggests fewer unnecessary elements, such as meaningless text or fictitious logos, thereby fulfilling essential requirements for motion graphics.
\end{itemize}
Overall, these results indicate that \ours{} is the most effective approach for generating motion graphics among all evaluated methods.

\section{Limitation} \label{sec:limit}
While our core idea, the integration of layer decomposition and script-based animation generation, shows solid quality in motion graphics generation, \ours{} has several limitations arising from the capability of layer decomposition. Although \ours{} generally animates text with high readability, small texts tend to have poor readability due to the difficulty of precisely segmenting tiny text. Accurately decomposing multiple overlapping layers is also challenging, resulting in unnatural object animations. Additionally, precise inpainting becomes difficult with large decomposed objects, leading to unnatural and visually unappealing results. 
\section{Conclusion}
This paper proposes the first end-to-end framework that generate motion graphics from a single raster image named \ours{}. 
\ours{} decomposes an input image into multiple HTML-represented layers using OCR, object detection, object segmentation, and image inpainting models, then generates an executable JavaScript animation script through an LMM.
It ensures active text motion with text readability and object motion with less distortion. 
Our experiments show that \ours{} outperforms state-of-the-art image-to-video methods for generating motion graphics.


{
    \small
    \bibliographystyle{ieeenat_fullname}
    \bibliography{main}
}

\end{document}